\documentclass[preprint,showpacs,preprintnumbers,amsmath,amssymb]{revtex4}

\usepackage{graphicx}
\usepackage{bm}
\usepackage{color}
\usepackage{amsmath}
\usepackage{amsfonts}
\usepackage{amssymb}
\usepackage{epstopdf}
\usepackage{ulem}

\begin{document}

\preprint{APS/123-QED}


\title{Magnetic Phase Diagram of the $S$=1/2 Triangular-Lattice Heisenberg Antiferromagnet Ba$_3$CoNb$_2$O$_9$}

\author{Kazuya Yokota, Nobuyuki Kurita, and Hidekazu Tanaka}
\affiliation{
Department of Physics, Tokyo Institute of Technology, Meguro-ku, Tokyo 152-8551, Japan
}
\date{\today}

\begin{abstract}

We report the results of low-temperature thermal and magnetic measurements on Ba$_3$CoNb$_2$O$_9$ powder,
described as a uniform triangular-lattice antiferromagnet (TLAF) with a fictitious spin-1/2. 
Ba$_3$CoNb$_2$O$_9$ is found to undergo two-step antiferromagnetic transitions 
at $T_{\rm N1}=1.39~{\rm K}$ and $T_{\rm N2}=1.13~{\rm K}$.
As the magnetic field is increased, both $T_{\rm N1}$ and $T_{\rm N2}$ monotonically decrease. 
The magnetic field vs temperature phase diagram indicates that the exchange interactions are 
nearly of the Heisenberg type with weak easy-axis anisotropy and that the exchange interaction 
between triangular lattices is crucial, in contrast to the case of the quasi-two-dimensional TLAF 
Ba$_3$CoSb$_2$O$_9$ [Susuki {\it et al}., Phys. Rev. Lett. \textbf{110}, 267201 (2013)].
\end{abstract}

\pacs{75.10.Jm, 75.40.Cx, 75.45.+j}

\maketitle


\section{Introduction}
Triangular-lattice antiferromagnets (TLAFs) have long been a subject of 
active research in condensed matter physics~\cite{Anderson_MRB1973,Mekata,Collins,Kawamura,Balents}. 
For classical Heisenberg spins, the antiferromagnetic nearest-neighbor exchange interaction produces triangular spin ordering. 
However, the classical ground state cannot be determined uniquely in a magnetic field 
because the ground state is continuously degenerate. 
In the presence of quantum fluctuations, the classical degeneracy can be lifted, 
so that a specific ground state is stabilized by the so-called order-by-disorder mechanism~\cite{Villian_JPhys1980}.
A two-dimensional (2D) $S\,{=}\,1/2$ triangular-lattice Heisenberg antiferromagnet (TLHAF) is one of the simplest quantum frustrated systems.
The combination of the reduced dimensionality, geometric frustration, 
and the smallest spin markedly enhances quantum fluctuations,
which can trigger exotic quantum states.
One macroscopic quantum manifestation predicted for 2D $S\,{=}\,1/2$ TLHAFs
is the stabilization of the ``up-up-down'' spin structure under an applied field
~\cite{Nishimori_JPSJ2013,Chubokov_CondMat1991,Farnell_JPCM2009,Honecker_JPCM1999,Sakai_PRB2011,Hotta}.
In the magnetization process, the quantum ground state appears as
a magnetization plateau at one-third of the saturation magnetization $M_{\rm s}$, 
whereas in the classical model, the magnetization increases monotonically with the magnetic field up to $M_{\rm s}$.

In spite of intensive research on 2D $S\,{=}\,1/2$ TLHAFs, details of the quantum effects 
including the $1/3$ magnetization plateau have not been elucidated definitively. 
This is mostly due to the lack of a model material that satisfactorily realizes a 2D $S\,{=}\,1/2$ TLHAF.
Thus far, most experimental studies on 2D $S=1/2$ TLHAFs in magnetic fields have 
focused on compounds comprising Cu$^{2+}$ magnetic ions such as 
Cs$_2$Cu$X_4$ ($X$\,=\,Cl, Br)~\cite{Ono_PRB2003,Fortune_PRL2009}.
However, the arrangement of Cu$X_4$ tetrahedra in Cs$_2$Cu$X_4$ does not have trigonal symmetry, 
which leads to a spatially anisotropic triangular lattice. 
The low-symmetric crystal lattice produces an antisymmetric interaction of the Dzyaloshinskiy-Moriya (DM) type. 
For these reasons, the magnetic model of Cs$_2$Cu$X_4$ becomes complicated.

\begin{figure}[thb]
\begin{center}
\vspace{2mm}
\includegraphics[width=0.6 \linewidth]{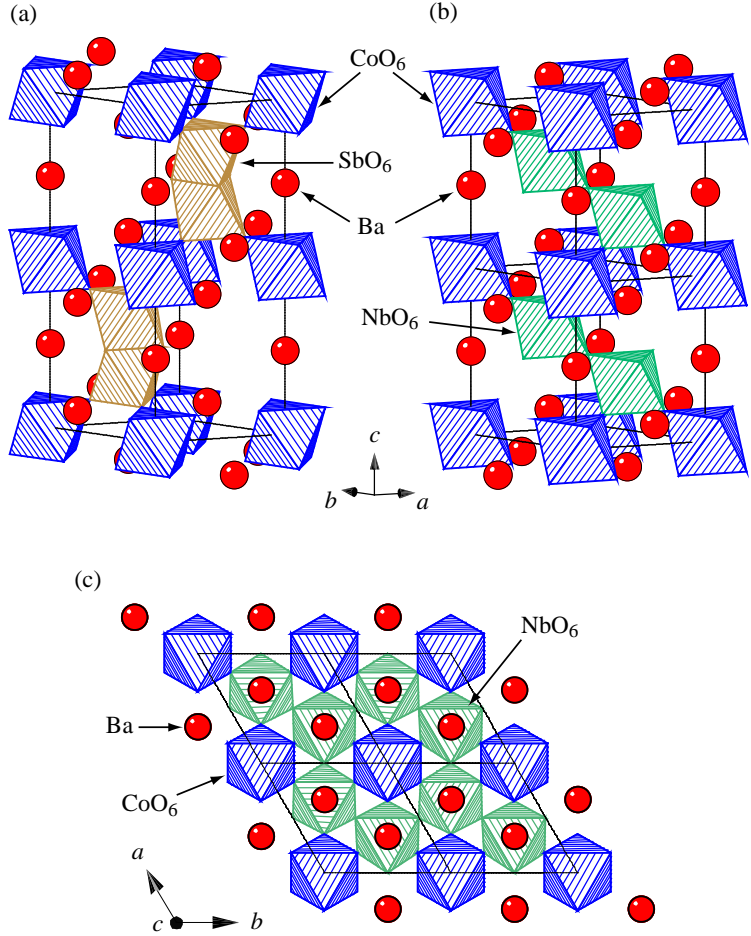}
\end{center}
\vspace{-2mm}
\caption{
(Color online) 
Schematic crystal structures of (a) Ba$_3$CoSb$_2$O$_9$ ($P6_3/mmc$) and (b) Ba$_3$CoNb$_2$O$_9$ ($P\bar{3}m1$). (c) Arrangement of CoO$_6$ and NbO$_6$ octahedra in Ba$_3$CoNb$_2$O$_9$ viewed along the $c$-axis. The solid lines show the chemical unit cell.
For both compounds, magnetic Co$^{2+}$ ions at the center of CoO$_6$ octahedra form a uniform triangular lattice in the $ab$ plane.
} 
\vspace{-2mm}
\label{struct}
\end{figure}

Recently, we have shown that Ba$_3$CoSb$_2$O$_9$ closely approximates 
an ideal 2D $S\,{=}\,1/2$ TLHAF~\cite{Shirata_PRL2012,Susuki_PRL2013}.
In Ba$_3$CoSb$_2$O$_9$, magnetic Co$^{2+}$ ions centered at CoO$_6$ octahedra form ``uniform" triangular lattice layers as shown in Fig.~\ref{struct}(a). 
Owing to the octahedral crystal field and spin-orbit interaction, the effective spin of Co$^{2+}$ can be represented by $S\,{=}\,1/2$ 
at low temperatures well below the spin-orbit coupling constant $\lambda/k_{\rm B}\,{\sim}\,250$\,K~\cite{Abragam,Lines,Oguchi}. Although crystal lattice of Ba$_3$CoSb$_2$O$_9$ is hexagonal, the crystal field is close to cubic field. This leads to the almost isotropic $g$-factor and exchange interaction between effective spins, unlike those in typical Co$^{2+}$ compounds~\cite{Shirata_PRL2012,Susuki_PRL2013}.
Actually, the electron paramagnetic resonance measurement revealed the almost isotropic $g$-factors of 3.84 and 3.87 for $H\,{\parallel}\,ab$ and $H\,{\parallel}\,c$, respectively~\cite{Susuki_PRL2013}.
 
When a magnetic field is applied parallel to the $ab$ plane, the quantum $1/3$ magnetization plateau is clearly observed~\cite{Susuki_PRL2013}.
Ba$_3$CoSb$_2$O$_9$ is the first example for which it was experimentally shown that the entire magnetization process including the quantum $1/3$ plateau 
exhibits good quantitative agreement with theoretical calculations.
In addition, we have found a possible high-field quantum state at approximately $3/5M_{\rm s}$ for $H \parallel ab$. 
To obtain a deeper understanding of the quantum phenomena in TLHAFs, it is worth investigating similar Co$^{2+}$-based materials 
and to compare the similarities and differences between their magnetic properties.

In this paper, we thus study the Nb-analog compound, Ba$_3$CoNb$_2$O$_9$,
through low-temperature magnetization and specific heat measurements.
Ba$_3$CoNb$_2$O$_9$ crystallizes in a hexagonal structure with the space group $P\bar{3}m1$ 
\,\cite{Treiber_ZAAC1982,Ting_JSSC2004,Ting}. 
Figures 1(b) and (c) show the crystal structure of Ba$_3$CoNb$_2$O$_9$.
The crystal structure of Ba$_3$CoNb$_2$O$_9$ differs from that of Ba$_3$CoSb$_2$O$_9$ 
in the stacking pattern of nonmagnetic $B$O$_6$ ($B$\,{=}\,Nb, Sb) octahedra.
Neighboring NbO$_6$ octahedra in Ba$_3$CoNb$_2$O$_9$ are linked sharing their corners, 
while for Ba$_3$CoSb$_2$O$_9$, neighboring SbO$_6$ octahedra are linked sharing their faces~\cite{Treiber_ZAAC1982,Ting,Doi_CondMat2004}.
Because of the highly symmetric crystal structure, the antisymmetric DM interaction is absent between the first-, second- and third-neighbor spins in the triangular lattice and between all spin pairs along the $c$ axis in both compounds.

\section{Experimental details}

Ba$_3$CoNb$_2$O$_9$ powder was prepared by the chemical reaction:
3BaCO$_3$ + CoO + Nb$_2$O$_5$ $\rightarrow$ Ba$_3$CoNb$_2$O$_9$ + 3CO$_{{2}}$. 
Reagent-grade materials were mixed in stoichiometric ratios and calcined 
at $1200\ {}^\circ\mathrm{C}$ for $20~{\rm h}$ in air. Ba$_3$CoNb$_2$O$_9$ was sintered at $1500\ {}^\circ\mathrm{C}$ for more than $20~{\rm h}$ after being pressed into a pellet. The brown powder samples obtained were confirmed to be Ba$_3$CoNb$_2$O$_9$ by X-ray diffraction. 

The specific heat was measured in the temperature range of 0.5 K to 300 K and under magnetic fields of up to 9 T
using a physical property measurement system (PPMS, Quantum Design) by the relaxation method. 
We also conducted magnetic measurements down to 0.4 K and up to $7~{\rm T}$ using a SQUID magnetometer (MPMS XL, Quantum Design) 
equipped with an iHelium3 option (IQUANTUM).

\section{Results}
\begin{figure}
\begin{center}
\includegraphics[width=0.6\linewidth]{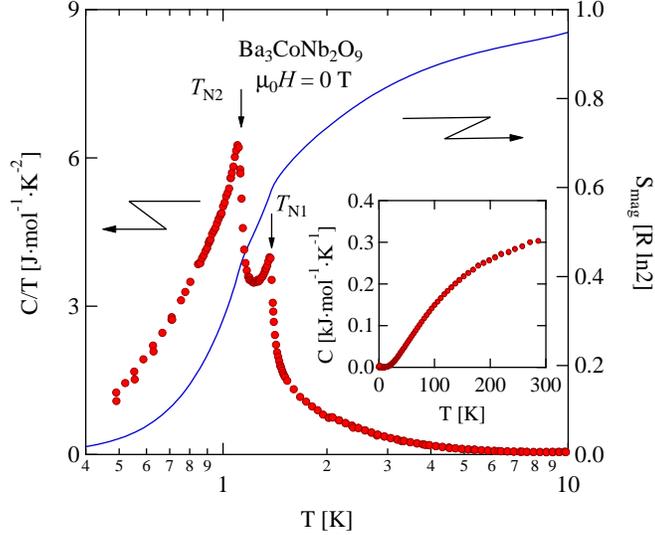}
\end{center}
\vspace{-4mm}
\caption{
(Color online) Temperature dependence of $C/T$ in Ba$_3$CoNb$_2$O$_9$ measured at zero magnetic field.
Magnetic phase transitions are observed at $T_{\rm N1}$=1.39 K and $T_{\rm N2}$=1.13 K, indicated by arrows. 
The solid curve represents the magnetic entropy $S_{\rm mag}$ in units of $R\ln{2}$.
The inset shows the specific heat below 300 K. 
} 
\vspace{-2mm}
\label{C_0T}
\end{figure}

The main panel of Fig.~\ref{C_0T} shows the temperature dependence of the specific heat in Ba$_3$CoNb$_2$O$_9$ 
divided by the temperature $C/T$ measured in the absence of a magnetic field. 
As the temperature decreases from 10 K, $C/T$ increases gradually and exhibits two pronounced peaks at $T_{\rm N1}$=1.39 K and  $T_{\rm N2}$=1.13 K 
indicative of successive magnetic phase transitions. 
These two phase transitions are also confirmed by the magnetization measurements shown below.

Because Ba$_3$CoNb$_2$O$_9$ is a magnetic insulator, the measured specific heat 
can be expressed as a sum of magnetic and phonon terms.
The high-temperature specific heat shown in the inset of Fig.~\ref{C_0T} is attributed to the phonon specific heat.
With decreasing temperature, $C/T$ at approximately 10 K becomes almost zero, 
in contrast to that at lower temperatures.
This indicates that the low-temperature specific heat below $\sim$10 K is mostly of magnetic origin, and the phonon contribution is negligible.
We thus evaluate the magnetic entropy $S_{\rm mag}$ of Ba$_3$CoNb$_2$O$_9$ 
by integrating the measured $C/T$ with respect to $T$ below 10 K~\cite{Smag}.
The entropy gain is estimated to be  $0.43~R\ln2$ at $T_{\rm N1}$ and $0.60~R\ln2$ at $T_{\rm N2}$.
With further increasing temperature, 
$S_{\rm mag}$ increases gradually and approaches approximately $R\ln2$ at 10 K.
This provides evidence that a pseudospin-1/2 description is indeed valid for Ba$_3$CoNb$_2$O$_9$
when $T < 10$ K.
Below $T_{\rm N2}$, the specific heat is proportional to $T^3$ with no residual term,
which indicates the 3D spin-wave dispersions in the ordered state.

\begin{figure}
\begin{center}
\includegraphics[width=0.6\linewidth]{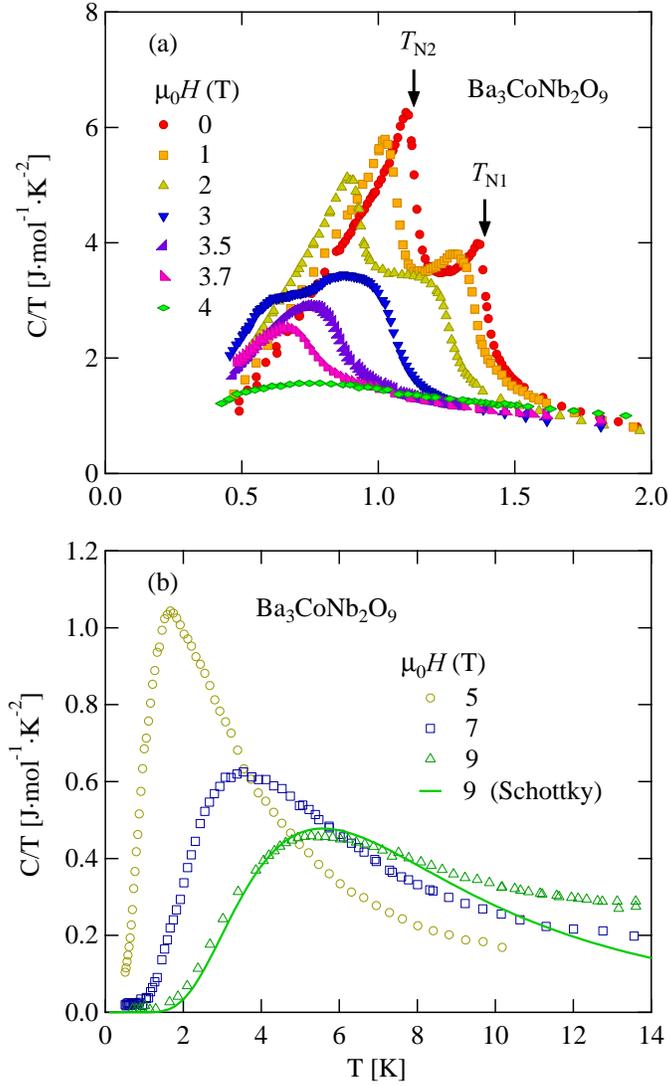}
\end{center}
\vspace{-4mm}
\caption{
(Color online) Temperature dependence of $C/T$ in Ba$_3$CoNb$_2$O$_9$ 
measured in magnetic fields of (a) $0-4~{\rm T}$ and (b) $5-9~{\rm T}$. 
The solid curve in (b) represents the Schottky contribution for the case of $H\,{=}\,9$ T~(see text).
} 
\vspace{-2mm}
\label{CvsT}
\end{figure}

Figure~\ref{CvsT}(a) shows $C/T$ vs $T$ measured in several magnetic fields of up to $4~{\rm T}$. 
With increasing magnetic field, the two peaks associated with $T_{\rm N1}$ and $T_{\rm N2}$ shift to lower temperatures and become smeared.
$T_{\rm N1}$ and $T_{\rm N2}$ are no more detectable for $H\,{>}\,3.7$ T and $H\,{>}\,3.0$ T, respectively.
At a higher field of 4 T, a broad maximum can be observed around $T_{\rm max}\,{=}\,0.7$ K. 
As shown in Fig.~\ref{CvsT}(b), $T_{\rm max}$ shifts to higher temperatures with increasing field,
while the peak height monotonically decreases.
A solid curve is fit to the $C/T$ data at 9 T 
using the following Schottky formula:
\begin{eqnarray}
C_{\rm Sch}=Nk_{\rm B}\left(\frac{g\mu_{\rm B}H}{2k_{\rm B}T}\right)^2\frac{1}{\cosh^2(g\mu_{\rm B}H/2k_{\rm B}T)},
\label{schottky}
\end{eqnarray}
where $N$, which is the number of atoms, and the $g$-factor are free parameters.
Equation (\ref{schottky}) expresses the Schottky specific heat due to the Zeeman splitting 
of an $S\,{=}\,1/2$ spin without exchange interaction between spins. 
The experimental data for the case of 9 T is well reproduced by eq.\,(\ref{schottky}) with $g\,{=}\,3.0$, 
which is in good agreement with the results obtained from the magnetization measurements shown below. 
The deviation of the fitting curve from the experimental data at high temperatures is ascribed to the phonon contribution. 
It is also found that although $T_{\rm max}$ and the peak height of the Schottky anomaly vary with the applied field, 
the magnetic entropy $S_{\rm mag}$ at 10 K is close to $R\ln2$, irrespective of the applied field. 
This is further evidence for the pseudospin-1/2 description of Ba$_3$CoNb$_2$O$_9$.

\begin{figure}
\begin{center}
\includegraphics[width=0.6\linewidth]{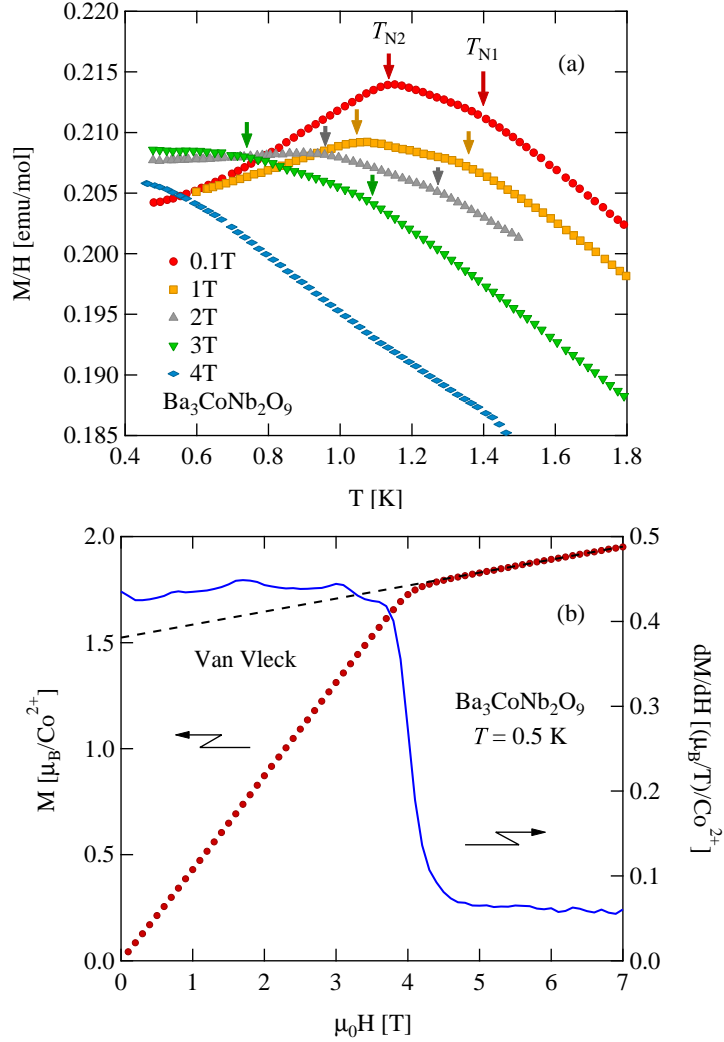}
\end{center}
\vspace{-4mm}
\caption{
(Color online) (a) Temperature dependence of magnetic susceptibility $M/H$ in Ba$_3$CoNb$_2$O$_9$ measured at several magnetic fields of up to 4 T, 
where the Van Vleck paramagnetic susceptibility has been subtracted. 
Arrows denote the ordering temperatures $T_{\rm N1}$ and $T_{\rm N2}$.
(b) Field dependence of raw magnetization $M_{\rm raw}$ (left) and 
its field derivative $dM_{\rm raw}/dH$ (right) in Ba$_3$CoNb$_2$O$_9$ measured at $0.5~{\rm K}$. 
The dashed line denotes the Van Vleck paramagnetism.
} 
\vspace{-2mm}
\label{chi}
\end{figure}

Figure~\ref{chi}(a) shows the $T$ dependence of the magnetization divided by the field $M/H$ in Ba$_3$CoNb$_2$O$_9$ 
measured at several magnetic fields of up to $4~{\rm T}$. 
The contribution of the Van Vleck paramagnetism is subtracted as will be discussed below. 
No thermal hysteresis is observed between the zero-field-cooled and field-cooled data. 
The two magnetization anomalies below 4 T, indicated by arrows for each set of data, are attributed to magnetic phase transitions. 
The transition temperatures $T_{\rm N1}$ and $T_{\rm N2}$ obtained at various magnetic fields are in good agreement with those obtained from specific heat measurements. 

Figure \ref{chi}(b) shows the raw magnetization $M_{\rm raw}$ in Ba$_3$CoNb$_2$O$_9$ and its field derivative as functions of $H$ measured at $0.5~{\rm K}$. 
The highest applied field of $7~{\rm T}$ was sufficient to reveal the entire magnetization process.
The saturation field is determined to be $H_{\rm s}\,{=}\,4.0$ T from the inflection point of the d$M_{\rm raw}$/d$H$ curve. 
This small saturation field indicates that the exchange interaction in Ba$_3$CoNb$_2$O$_9$ is much smaller than that in Ba$_3$CoSb$_2$O$_9$, 
in which the saturation field is $H_{\rm s}\,{\simeq}\,32$ T~\cite{Shirata_PRL2012,Susuki_PRL2013}.
Above $H_{\rm s}$, $M_{\rm raw}$ increases linearly with increasing field because of the large $T$-independent Van Vleck paramagnetism 
characteristic of the Co$^{2+}$ ion in the octahedral environment~\cite{Oguchi}, as observed in Ba$_3$CoSb$_2$O$_9$~\cite{Shirata_PRL2012,Susuki_PRL2013}.
From the magnetization slope above $H_{\rm s}$, we evaluate the Van Vleck paramagnetic susceptibility to be
$\chi_{\rm VV}\,{=}\,6.1 \times 10^{-2}~\mu_{\rm B}/({\rm Co}^{2+}{\rm T})\,{=}\,3.4\,{\times}\,10^{-2}~{\rm emu/mol}$. 
By subtracting the Van Vleck term, the saturation magnetization is obtained to be $M_{\rm s}\,{=}\,1.5~{\mu_{\rm B}/{\rm Co}^{2+}}$, 
which gives an average $g$-factor of 3.0.

As seen from Fig.\,\ref{chi}(b), the magnetization anomaly at $H_{\rm s}$ is considerably sharp in spite of powdered sample. If the $g$-factor and/or exchange interaction are anisotropic, the saturation field depend strongly on the field direction, so that the magnetization anomaly at $H_{\rm s}$ for powdered sample should be smeared. The sharp magnetization anomaly at $H_{\rm s}$ observed in Ba$_3$CoNb$_2$O$_9$ demonstrates that both of the $g$-factor and exchange interaction are nearly isotropic, as in the case for the Sb-analog compound~\cite{Susuki_PRL2013}.

\begin{figure}
\begin{center}
\includegraphics[width=0.5\linewidth]{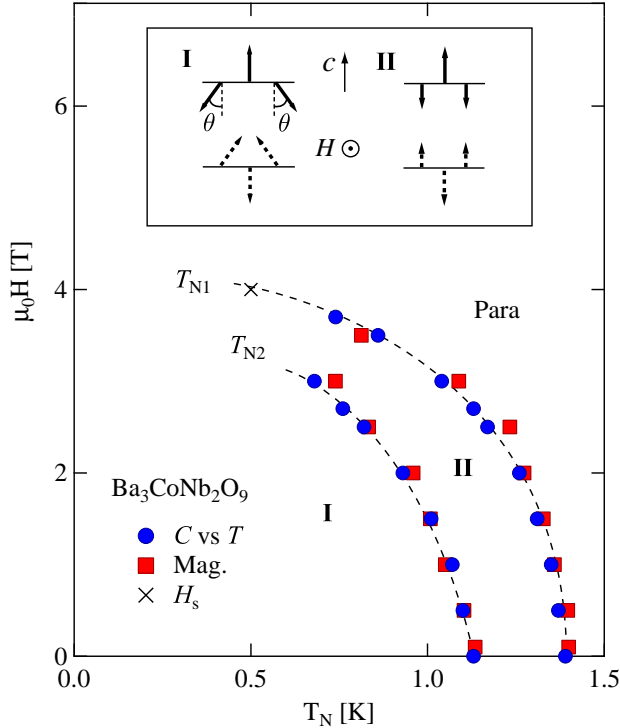}
\end{center}
\vspace{-4mm}
\caption{
(Color online) Magnetic field vs temperature phase diagram of Ba$_3$CoNb$_2$O$_9$ powder obtained from the present specific heat and magnetization measurements. 
Dashed curves are guides to the eyes.
The spin configurations expected in the two ordered phases I and II are displayed in the inset. 
Solid and dashed arrows denote the spin configurations on the two neighboring triangular layers. 
The magnetic field is assumed to be perpendicular to the $c$ axis.
} 
\vspace{-2mm}
\label{HT}
\end{figure}

Figure~\ref{HT} shows the magnetic phase diagram of Ba$_3$CoNb$_2$O$_9$,
determined via specific heat and magnetization measurements. 
It has been theoretically shown that two-step magnetic ordering occurs in TLAFs when the magnetic anisotropy is of the easy-axis type, 
while the ordering is single-step when the anisotropy is of the easy-plane type~\cite{Matsubara_JPSJ1982,Miyashita_JPSJ1985}. 
Thus, we infer that the successive phase transitions observed in Ba$_3$CoNb$_2$O$_9$ arise from easy-axis magnetic anisotropy. 
Note that the magnetic anisotropy in Ba$_3$CoSb$_2$O$_9$ is of the easy-plane type~\cite{Susuki_PRL2013}. 
A small difference in the trigonal crystal field gives rise to different signs of magnetic anisotropy in these two compounds.

For the powder sample with hexagonal symmetry, the physical anomaly for the case of $H\,{\perp}\,c$ is more pronounced than that for $H\,{\parallel}\,c$ 
because the probability of $H\,{\perp}\,c$ is twice as large as that of $H\,{\parallel}\,c$. 
This was observed in the magnetization process of Ba$_3$CoSb$_2$O$_9$ powder~\cite{Shirata_PRL2012,Susuki_PRL2013}. 
Thus, we deduce that the phase diagram shown in Fig.~\ref{HT} approximates the phase diagram for $H\,{\perp}\,c$ in Ba$_3$CoNb$_2$O$_9$. 
This phase diagram is considerably different from that for a quasi-2D TLAF with easy-axis anisotropy, as observed in Rb$_4$Mn(MoO$_4$)$_3$~\cite{Ishii}, 
but similar to that for a 3D TLAF with easy-axis anisotropy, 
in which the antiferromagnetic interlayer exchange interaction is of the same order of magnitude as the intralayer exchange interaction, 
as discussed theoretically by Plumer {\it et al.}~\cite{Plumer_PRL1988,Plumer_PRB1989}. 
Note that when the antiferromagnetic interlayer exchange interaction is much larger than the intralayer exchange interaction, 
both $T_{\rm N1}$ and $T_{\rm N2}$ increase with increasing magnetic field, as observed in CsNiCl$_3$~\cite{Johnson}. 
For quasi-2D TLAF with easy-axis anisotropy, the temperature range of the intermediate phase becomes narrow 
and vanishes with increasing magnetic field for $H\,{\perp}\,c$\,\cite{Ishii}. 
Thus, we infer that Ba$_3$CoNb$_2$O$_9$ can be magnetically described as an antiferromagnetically stacked 3D TLAF with easy-axis anisotropy. 
The reason why the intralayer exchange interaction is so small as compared with that in Ba$_3$CoSb$_2$O$_9$ will be discussed later.

Possible spin configurations at each magnetic phase of Ba$_3$CoNb$_2$O$_9$ are illustrated by arrows in Fig.\,\ref{HT}. 
With decreasing temperature, the $c$ axis component of the spin becomes ordered at $T_{\rm N1}$ with the ferrimagnetic structure in the $ab$ plane 
as $\langle S^z_1\rangle\,{<}\,{-}\,2\langle S^z_2\rangle\,{=}\,{-}\,2\langle S^z_3\rangle$. 
At $T_{\rm N2}$, the $ab$ plane component of the spin is ordered, such that the spins form a triangular structure in the plane including the $c$ axis, 
i.e., one-third of the spins are parallel to the $c$ axis and the remainder are canted away from the $c$ axis. 
The spins on the neighboring layers are aligned antiparallel to each other 
because of the antiferromagnetic interlayer exchange interaction, as shown in Fig.\,\ref{HT}. 
It is known that the temperature range of the intermediate phase $(T_{\rm N1}\,{-}\,T_{\rm N2})/T_{\rm N1}$ is a measure 
of the magnitude of the easy-axis anisotropy relative to the intralayer exchange interaction~\cite{Matsubara_JPSJ1982,Miyashita_JPSJ1985}. 
The narrow intermediate phase observed in Ba$_3$CoNb$_2$O$_9$ indicates that the easy-axis anisotropy is considerably smaller than the intralayer interaction.

In Ba$_3$CoSb$_2$O$_9$, in which the antiferromagnetic intralayer exchange interaction is much larger than the interlayer exchange interaction, 
the quantum $1/3$ magnetization plateau was clearly observed in both the powder~\cite{Shirata_PRL2012} and single crystals~\cite{Susuki_PRL2013}, 
while no such feature was detectable in Ba$_3$CoNb$_2$O$_9$ powder samples.
When the antiferromagnetic interlayer exchange interaction is comparable to the intralayer exchange interaction, 
the spin components perpendicular to the magnetic field form a triangular structure as shown in Fig.\,\ref{HT}. 
With increasing magnetic field, the spin component parallel to the magnetic field increases and 
the canting angle $\theta$ decreases to become zero at a finite temperature. 
This leads to a transition from the triangular state to the collinear state~\cite{Plumer_PRB1989}.

\begin{figure}
\begin{center}
\includegraphics[width=0.5\linewidth]{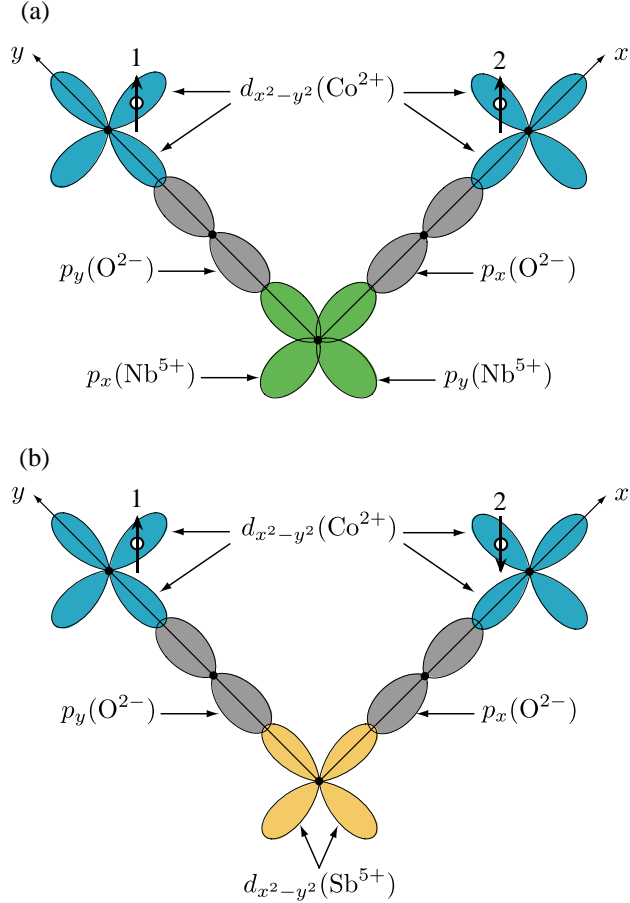}
\end{center}
\vspace{-2mm}
\caption{
(Color online) Illustrations of orbital configurations related to superexchange interactions in Ba$_3$Co$M_2$O$_9$ through Co$^{2+}-$\,O$^{2-}$\,$-$\,$M^{5+}-$\,O$^{2-}$\,$-$\,Co$^{2+}$ 
for (a) $M$\,{=}\,Nb and (b) Sb.
} 
\vspace{-2mm}
\label{Orbital}
\end{figure}

Next, we discuss the reason why the antiferromagnetic intralayer exchange interaction in Ba$_3$CoNb$_2$O$_9$ 
is much smaller than that in Ba$_3$CoSb$_2$O$_9$, on the basis of the Kanamori theory.\cite{Kanamori}
In Ref.\,\onlinecite{Kanamori}, it has been theoretically shown that, when the magnetic cation is subjected to the octahedral crystalline field, the sign of the superexchange interaction via nonmagnetic anion is closely connected with the orbital states of cation and anion. Kanamori theory is useful to discuss qualitatively whether the superexchange interaction is ferromagnetic or antiferromagnetic.
In Ba$_3$CoNb$_2$O$_9$ and Ba$_3$CoSb$_2$O$_9$, Co$^{2+}$ spins in the same layer interact via superexchange interactions 
through Co$^{2+}-$\,O$^{2-}-$\,O$^{2-}-$\,Co$^{2+}$ and Co$^{2+}-$\,O$^{2-}-$\,$M^{5+}-$\,O$^{2-}-$\,Co$^{2+}$ paths with $M$\,=\,Nb, Sb. 
The superexchange interaction through Co$^{2+}-$\,O$^{2-}-$\,O$^{2-}-$\,Co$^{2+}$ is common to both systems and should be antiferromagnetic as observed in many magnetic compounds.
Thus, it is considered that the difference in the exchange interaction between these two systems arises from the filled outermost orbitals, 
which are $4p$ and $4d$ orbitals for Ba$_3$CoNb$_2$O$_9$ and Ba$_3$CoSb$_2$O$_9$, respectively. 
We consider the superexchange interaction between hole spins on the $d_{x^2{-}y^2}$ orbitals of Co$^{2+}$ ions. 
The possible exchange paths are illustrated in Fig.~\ref{Orbital}. Here, we assume for simplification that successive Co$^{2+}$, O$^{2-}$ and M$^{5+}$ ions are on a straight line and the bond angle of O$^{2-}$\,$-$\,M$^{5+}$\,$-$\,O$^{2-}$ is 90$^{\circ}$.
The superexchange mechanism for $M$\,=\,Nb is based on the following perturbation process: 
(1)~Hole 1 with up spin on the left Co$^{2+}$ is first transferred to the $p_y$ orbital of O$^{2-}$, 
which is combined with the $p_y$ orbital of Nb$^{5+}$ to form a molecular orbital. 
(2)~Hole 2 on the right Co$^{2+}$ is also transferred to the other molecular orbital composed of two $p_x$ orbitals of O$^{2-}$ and Nb$^{5+}$. 
In this case, the total energy of the system is decreased when the spin of hole 2 is up, because in the Nb$^{5+}$ ion, 
the two hole spins on the $p_y$ and $p_x$ orbitals have to be parallel owing to the Hund rule. 
(3)~The two holes are transferred back to the $d_{x^2{-}y^2}$ orbitals of Co$^{2+}$. 
Consequently, a ferromagnetic superexchange interaction takes place.
On the other hand, the superexchange for $M$\,=\,Sb should be antiferromagnetic because in the Sb$^{5+}$ ion, 
the two hole spins on the same $d_{x^2{-}y^2}$ have to be antiparallel to each other owing to the Pauli principle. 
Similar results are also obtained for the hole spins on the $d_{3z^2{-}r^2}$ orbitals of Co$^{2+}$ ions. 
We infer that for $M$\,=\,Nb, antiferromagnetic and ferromagnetic superexchange interactions 
through Co$^{2+}-$\,O$^{2-}-$\,O$^{2-}-$\,Co$^{2+}$ and Co$^{2+}-$\,O$^{2-}-$\,$M^{5+}-$\,O$^{2-}-$\,Co$^{2+}$, respectively, mostly cancel out, 
so that the resultant antiferromagnetic exchange interaction becomes small. 
However, for $M$\,=\,Sb, the superexchange interactions through these two paths are both antiferromagnetic, 
and the total antiferromagnetic exchange interaction is enhanced. 
For this reason, the antiferromagnetic intralayer exchange interaction 
in Ba$_3$CoNb$_2$O$_9$ could be much smaller than that in Ba$_3$CoSb$_2$O$_9$.
In order to determine the detailed magnetic parameters, further experimental and theoretical approaches such as a neutron scattering measurement and density-functional calculations are needed.

\section{Conclusion}
In summary, we have performed low-temperature specific heat and magnetization measurements on Ba$_3$CoNb$_2$O$_9$ powder. 
Ba$_3$CoNb$_2$O$_9$ exhibits two magnetic transitions at $T_{\rm N1}\,{=}\,1.39$ K and $T_{\rm N2}\,{=}\,1.13~{\rm K}$ with a narrow intermediate phase, 
which arise from the weak easy-axis magnetic anisotropy. 
The magnetization saturates at $H_{\rm s}\,{=}\,4.0$ T. 
This small saturation field shows that the intralayer exchange interaction is much smaller than that in Ba$_3$CoSb$_2$O$_9$. 
The difference in the filled outermost orbitals of Nb$^{5+}$ and Sb$^{5+}$ gives rise to considerably different exchange interactions in both systems. 
The magnetic field vs temperature phase diagram shown in Fig.\,\ref{HT}, which is considered to approximate that for $H\,{\perp}\,c$, 
is in accordance with that for a TLAF with a strong interlayer exchange interaction and small easy-axis anisotropy~\cite{Plumer_PRL1988,Plumer_PRB1989}. Therefore, Ba$_3$CoNb$_2$O$_9$ can be described as an $S\,{=}\,1/2$ antiferromagnetically stacked 3D TLAF with easy-axis anisotropy,
in contrast to quasi-2D Ba$_3$CoSb$_2$O$_9$~\cite{Shirata_PRL2012,Susuki_PRL2013}. 
Note that our conclusion is different from the conclusion by Lee {\it et al.}~\cite{Lee}, although most of our experimental results are consistent with theirs. They attributed the field-induced phase transitions observed in Ba$_3$CoNb$_2$O$_9$ to quantum fluctuations characteristic of quasi-2D TLAF.
Ba$_3$CoNb$_2$O$_9$ could be a useful model for obtaining a comprehensive understanding of the $S\,{=}\,1/2$ TLAFs.

\section*{ACKNOWLEDGMENTS}
This work was supported by a Grant-in-Aid for Scientific Research (A) from the Japan Society for the Promotion of Science, and the Global COE Program 
``Nanoscience and Quantum Physics'' at Tokyo Tech. funded by the Ministry of Education, Culture, Sports, Science and Technology of Japan.


\begin{thebibliography}{25}
\bibitem{Smag}
The temperature dependence of $S_{\rm mag}$ for $T \le$ 0.4 K is deduced 
from a smooth extrapolation using the higher temperature $C/T$ data.
\end{thebibliography}
\end{document}